\documentclass[twocolumn,aps,showpacs]{revtex4}
\usepackage{epsfig,color}

\begin{document}
%\draft
\title{How Sensitive are Di-Leptons from $\rho$ Mesons to the High
Baryon Density Region?}

\author{S.Vogel ${}^{1}$, H. Petersen${}^{1,2}$, K. Schmidt${}^{1}$, E. Santini${}^{1}$, C. Sturm${}^{1}$
J. Aichelin${}^{3}$ and M. Bleicher${}^{1}$}

\address{\vspace*{.5cm}${}^{1}$Johann Wolfgang Goethe-Universit\"at, \\
Max-von-Laue-Stra\ss{}e 1,\\
60438 Frankfurt am Main, Germany}

\address{\vspace*{.3cm}${}^{2}$Frankfurt Institute for Advanced Studies (FIAS), \\
Ruth-Moufang-Stra\ss{}e 1,\\
60438 Frankfurt am Main, Germany}

\address{\vspace*{.3cm}${}^{3}$SUBATECH,
Laboratoire de Physique Subatomique et des Technologies Associ\'ees \\
University of Nantes - IN2P3/CNRS - Ecole des Mines de Nantes \\
4 rue Alfred Kastler, F-44072 Nantes Cedex 03, France}

\begin{abstract}
We show that the measurement of di-leptons might provide only a restricted view into the most dense stages 
of heavy ion reactions. Thus, possible studies of meson and baryon properties at high baryon densities, as e.g. done 
at GSI-HADES and envisioned for FAIR-CBM, might observe weaker effects than currently expected in certain approaches.  
We argue that the strong absorption of resonances in the high baryon density region of the heavy ion collision 
masks information from the early hot and dense phase due to a strong increase of the total decay width because 
of collisional broadening. To obtain additional information, we also compare the currently used approaches to extract
di-leptons from transport simulations - i.e. shining, only vector mesons from final baryon resonance decays and 
instant emission of di-leptons and find a strong sensitivity on the method employed in particular at FAIR 
and SPS energies. It is shown explicitly that a restriction to $\rho$ meson (and therefore di-lepton) production 
only in final state baryon resonance decays provide a strong bias towards rather low baryon densities. 
The results presented are obtained from UrQMD v2.3 calculations using the standard set-up.
\vspace{.6cm}
\end{abstract}


\maketitle

Quantum-Chromo-Dynamics (QCD) predicts that the properties of hadrons change when they are
brought into a (hot and/or dense) nuclear environment \cite{Hatsuda:1991ez,Brown:1991kk}. This modification is 
due to the interaction with the surrounding medium which eventually leads to chiral 
symmetry restoration at high baryon densities and/or high temperatures  \cite{Rapp:1999ej}. 
The experimental verification of this theoretical prediction is one of the most challenging questions
in modern strongly interacting matter physics.

Among the non-strange mesons the $\rho$ meson plays a dominant role
in these investigations. It has a short lifetime and therefore it
has a large probability to decay inside the reaction zone when created
in heavy ion collisions. It couples strongly to nuclear resonances
and, most important, it has a non-negligible chance to decay into
di-leptons which leave the interaction zone essentially without any
further interaction. Thus, the di-lepton channel seems to offer
a unique chance to study the high baryon density properties of the $\rho$
meson.

Theoretically the question of how the spectral function of the $\rho$ meson changes in the medium is 
still under active discussion. There is certain theoretical evidence that the $\rho$ meson
is broadened if put into the nuclear medium \cite{Pisarski:1995xu,Rapp:1995zy,Rapp:1997fs}. 
In contrast, Hatsuda and Lee predicted a lowering of the $\rho$ meson mass
in a nuclear environment  based on QCD sum rules calculations \cite{Hatsuda:1991ez}. A result which has also been
found by Brown and Rho \cite{Brown:1991kk,Li:1995qm,Brown:2001nh}. On the other hand, more recent
calculations indicate that the pole mass of the $\rho$ meson remains almost unchanged in the nuclear 
medium \cite{vanHees:2006ng,Ruppert:2006hf,Dusling:2006yv}. 
However, these calculations rely on specific assumptions on the coupling strength of the
$\rho$ meson to the nuclear resonances and on the branching ratios whose validity can presently only be 
proven by comparison to experimental data. For the present status of the theoretical spectral
function calculations for vector mesons we refer to
\cite{nucl-th/0108017,nucl-th/0607061,Gallmeister:2006yt,Rapp:2006rh,Santini:2008pk}. 

Experiments have been launched to verify these theoretical
predictions. In proton-nucleus collisions \cite{Naruki:2005kd,Ozawa:2000iw} at 12 GeV a decrease of the
$\rho$ meson mass with increasing baryon density $\rho_B$ as $m(\rho_B)/m(0) = 1 - 0.09 \rho_B /\rho_{B0}$ 
- about half of the value predicted by theory - but without an increase of the $\rho$ meson's width has been
reported. The CLAS collaboration reports that the experimental data of photon-induced reactions is 
compatible with no shift of the $\rho$ meson pole mass and no additional broadening to the theoretically estimated 
collisional broadening \cite{Weygand:2007ky}. In contrast, the di-lepton data in In+In collisions at 158 AGeV
\cite{Arnaldi:2006jq} are best described using essentially the free $\rho$ meson pole mass but a
considerable broadening of the spectral function. At lower energies of 2 AGeV 
(see \cite{Agakishiev:2007ts} for the newest data at 1~GeV), 
the HADES collaboration has recently published di-lepton spectra \cite{Agakichiev:2006tg}. 
Here, a deviation from the yield calculated from a hadronic cocktail fit in the vicinity
of the $\rho$ meson mass is visible but due to the many sources of di-leptons there is no conclusive explanation 
for that suppression yet. How much of these different experimental findings can be exclusively attributed to the
different environments, i.e. cold nuclear matter in proton-nucleus reactions, an
expanding meson dominated fireball after a possible phase transition
from a quark gluon plasma in high energy nucleus-nucleus collisions at the SPS or a baryon dominated
expansion in reactions at about 2 AGeV is still a matter of debate. 

To link the final state di-lepton data to the in-medium spectral functions of the hadrons detailed 
quantitative theoretical simulations of the baryon density distribution at the $\rho$ meson 
production and decay/absorption 
point are necessary. This allows then to calculate di-lepton spectra from the simulations. Unfortunately, up to now
different approaches are used to convert the calculated hadron spectra into di-leptons.
In general four different approaches can be identified:

\begin{itemize}
\item Explicit propagation of stable particles, baryon and meson resonances, decays of resonances into other 
mesons (especially baryonic resonances into $\rho$ mesons), as well as $\pi\pi \rightarrow \rho$ scattering. 
Di-leptons are emitted continuously from vector mesons and baryon resonances with their respective locally given total 
width while the resonances are propagated 
(``shining'') \cite{Koch:1996wy,Ernst:1997yy,Cassing:1999es,Cassing:2000bj,Bratkovskaya:2007jk},

\item Explicit propagation of stable particles, baryon and meson resonances, decay of resonances into other 
mesons (especially baryonic resonances into $\rho$ mesons), as well as $\pi\pi \rightarrow \rho$ scattering.
Di-leptons are only emitted at the point of decay (not absorption) of the hadronic resonance \cite{Schumacher:2006wc},

\item Explicit propagation of nucleons, pions and nucleon resonances while vector meson degrees of freedom 
are not propagated explicitly. Di-leptons are produced via the eVMD model from nucleonic resonances at the 
point of the decay of the nucleonic resonances. This model assumes di-lepton production via 
intermediate $\rho / \omega$ states  \cite{Cozma:2006vp,Santini:2008pk},

\item Explicit propagation of nucleons, pions, kaons and $\Delta$ resonance. Di-leptons are produced by 
decay of nucleonic resonances according to the branching ratios. This is supplemented by the di-lepton production 
from mesonic resonances, which are calculated by folding the $\sqrt{s}$ distribution of the elementary 
nucleon-nucleon collisions with the mesonic production cross sections and the corresponding branching 
ratios into di-leptons \cite{Thomere:2007cj}.
\end{itemize}
\begin{figure}[hbt]
\vspace*{-.5cm} %\centerline{\resizebox*{!}{0.5\textheight}
\epsfig{file=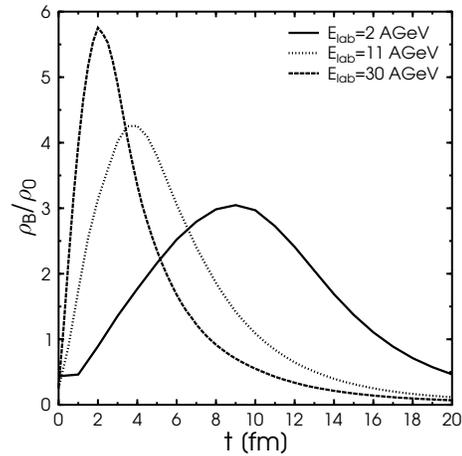,width=7.5cm} \caption {(Color online) Time evolution of the local rest 
frame baryon density $\rho_B$ averaged over the positions of the individual hadrons for central Au+Au/Pb+Pb 
reactions at various beam energies.}
\label{rhob}
\end{figure}

While the first method is sensitive to all stages of the collision and allows for a dynamical 
treatment of the collisional broadening, the other methods rely on the actual 
decay of the meson or baryon resonance. Thus, the different approaches probe different baryon density 
regimes and might therefore provide  different results for the extracted di-lepton rates. 
Another problem is posed by the implementation of 
bremsstrahlung especially at low beam energies. This discussion has recently been revived by
the calculations of \cite{Shyam:2003cn,Kaptari:2005qz,Thomere:2007cj,Bratkovskaya:2007jk}. 

In this situation it is necessary to study the general differences between the above discussed approaches and
explore the baryon density probed by the $\rho$ meson in the FAIR energy regime. 
This helps to provide a theoretical error margin for further detailed model studies on the change of the 
in-medium spectral functions at these energies. 

We perform this study for massive nuclear reactions in the energy range of  $2A~{\rm GeV}\le E_{\rm lab}\le 30A$~GeV. 
This range marks the expected transition towards chiral symmetry restoration, but also the transition from
baryon dominated to meson dominated matter. Dedicated facilities to explore this energy domain are the 
FAIR project at GSI \cite{Augustin:2008gy} and the critRHIC program at BNL \cite{Stephans:2006tg}. 
For our studies we employ the UrQMD(v2.3) 
model \cite{Bass:1998ca,Bleicher:1999xi,Petersen:2008kb}. 
It is a non-equilibrium transport approach based on the
covariant propagation of hadrons and strings. All elementary cross sections are
calculated by the principle of detailed balance, from the additive quark model or are derived from
available data. This model allows to study the full space time
evolution for all hadrons, resonances and their decay products. It
permits to explore the emission patterns of the resonances in detail
and to gain insight into the origin of the resonances
\cite{Bleicher:2002dm,Bleicher:2002rx,Bleicher:2003ij,Vogel:2006kd,Vogel:2005pd}. 

In this approach the
resonances are treated with their vacuum properties during the whole
evolution of the reaction. However, one should note that the particle properties are dynamically 
modified in a hot and/or dense medium due to the coupling of the $\rho$ meson to the surrounding hadrons (at SIS energies 
especially the baryon resonances are important \cite{Vogel:2005pd}). 
The $\rho$ meson is assumed to have a lifetime according to an exponential distribution with a mean lifetime $\tau$ 
of $1/\Gamma_{pole} \simeq 1/150$~MeV, in addition collisional broadening is implicitly taken
into account for the calculation of di-leptons by the interaction of the  $\rho$ meson with the evolving medium. 
\begin{figure}[t]
\vspace*{-.5cm} %\centerline{\resizebox*{!}{0.5\textheight}
\epsfig{file=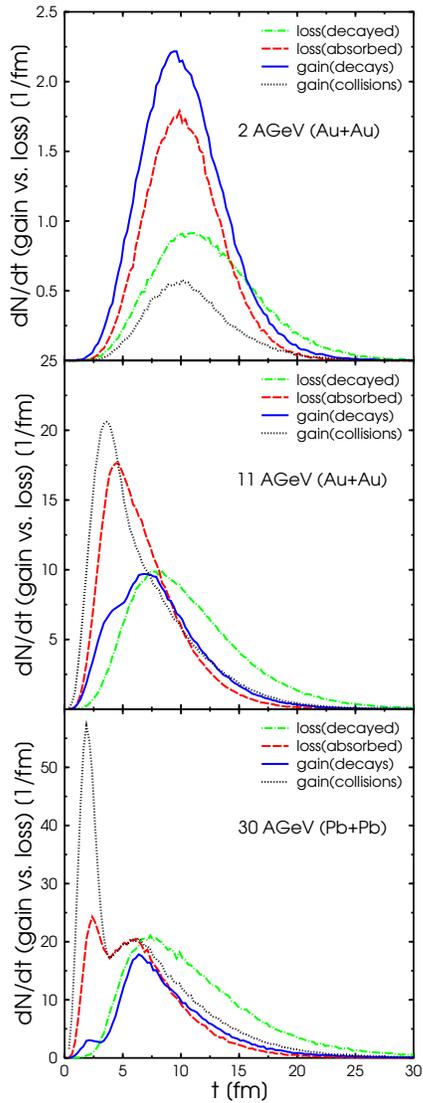,width=6cm} \caption {(Color online) Gain and loss rates
of $\rho$ mesons separated gain from collisions (``gain(collisions)''), gain from decay (``gain(decays)'') and loss due to absorption (``loss(absorbed)'') and loss in decays (``loss(decayed)''). From top to
bottom we display central Au+Au/Pb+Pb collisions at 2, 11 and 30 AGeV. }
\label{rate}
\end{figure}
\begin{figure}[t]
\vspace*{-.5cm}
%\centerline{\resizebox*{!}{0.5\textheight}}
\epsfig{file=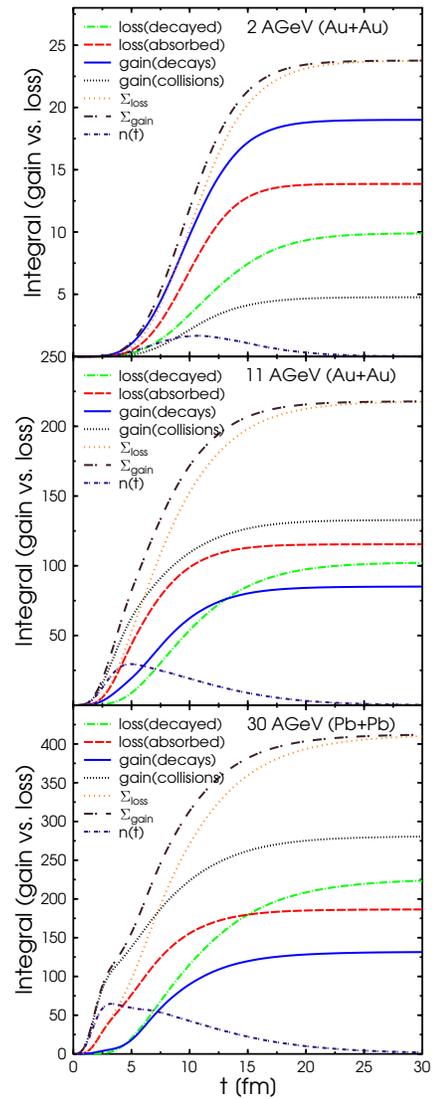,width=6cm}
\caption{(Color online) Gain and loss of
$\rho$ mesons separated for collisions and decay. From top to bottom
we display central Au+Au/Pb+Pb collisions at 2, 11 and 30 AGeV. We show as
well the difference of gain and loss, the number of $\rho$ mesons present
in the system as a function of time. From top to bottom we display
central Au+Au/Pb+Pb collisions at 2, 11 and 30 AGeV.}
\label{number}
\end{figure}

UrQMD has been successfully applied to study light and heavy ion reactions from
$E_{\rm beam}= 90A$~MeV to $\sqrt{s_{NN}}=200$ GeV. The present version of the transport
model is able to describe the yields and the transverse momentum spectra of
different particles in proton-proton, proton-nucleus and nucleus-nucleus collisions
\cite{Bratkovskaya:2004kv,Petersen:2008kb}. Detailed comparisons of UrQMD with a
large body of experimental data at SIS and FAIR energies can be found in
\cite{Sturm:2000dm,Petersen:2008kb}. For further details of the model the reader is
referred to \cite{Bass:1998ca,Bleicher:1999xi,Petersen:2008kb}.

Let us start the discussion by displaying the time evolution of the baryon density in central Au+Au/Pb+Pb 
reactions at 2, 11 and 30A GeV (see Fig. \ref{rhob}). The baryon density is averaged over all hadron positions 
and is calculated locally in the rest frame of the baryon current (Eckart frame) averaged over the 
position of every baryon as $\rho_B=j^0$ with $j^\mu = (\rho_B,\vec{0})$. Details on the calculation of the baryon density
are discussed in the Appendix. 
Note that the maximal baryon density grows with increasing beam energy. The question to be asked is, 
how sensitive are di-lepton observables on the high baryon density stage of the collisions?

In order to investigate this question, we first review the different production and loss mechanisms for 
the $\rho$ meson. In UrQMD the $\rho$ meson can be produced from the decay of a high mass (meson or baryon) 
resonance or directly in a collision of two particles (e.g. $\pi+\pi \rightarrow \rho$) which includes also 
the production from string fragmentation. The $\rho$ meson is destroyed by two different mechanisms. It can
 decay (``loss (decayed)'', e.g. $\rho \rightarrow \pi^+\pi^-$) or it can be absorbed in 
collisions (``loss (absorbed)'', e.g. $\rho + \pi \rightarrow a_1$).
\begin{figure}[t]
\vspace*{-.5cm} %\centerline{\resizebox*{!}{0.5\textheight}
\epsfig{file=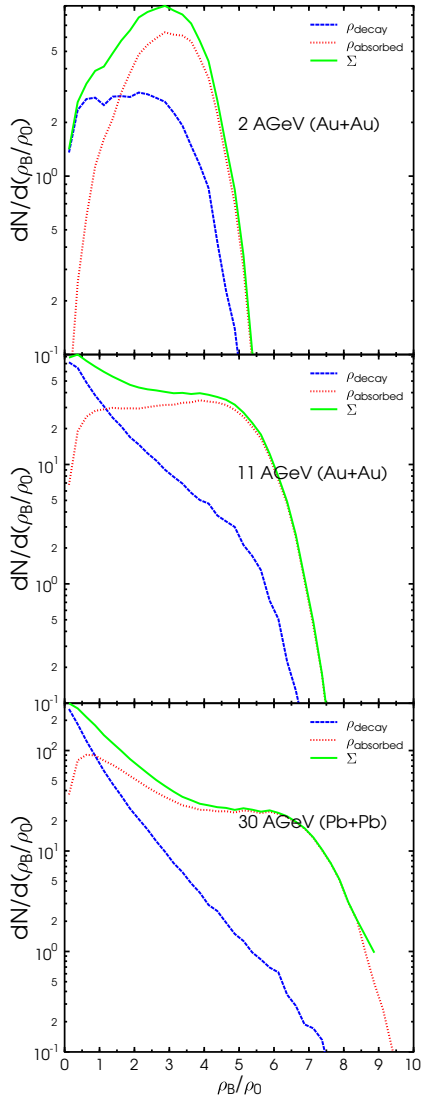,width=6cm}
\caption{(Color online) Baryon density distribution
at the space points where the $\rho$ mesons decay. From top to
bottom we display central Au+Au/Pb+Pb collisions at 2, 11 and 30 AGeV}
\label{dens}
\end{figure}
\begin{figure}[t]
\vspace*{-.5cm} %\centerline{\resizebox*{!}{0.5\textheight}
\epsfig{file=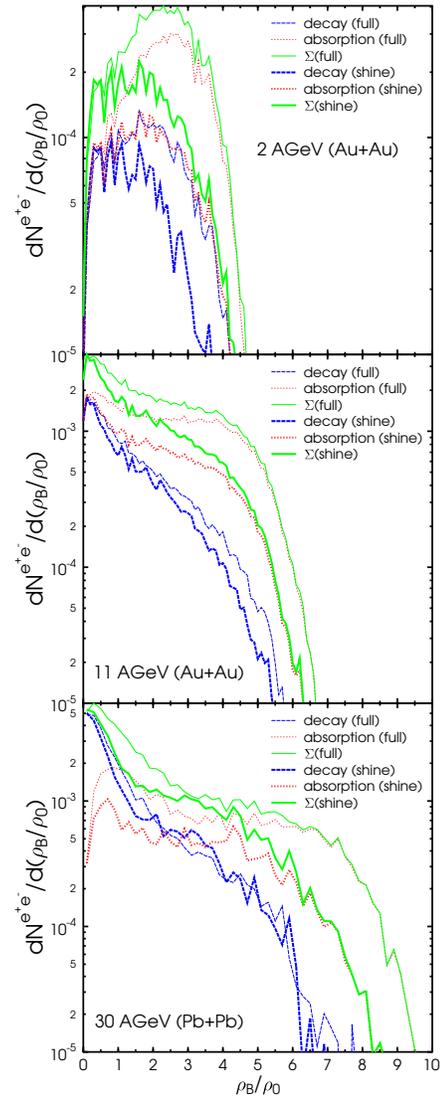,width=6cm}
\caption{(Color online) Distribution of the baryon density at which the $e^+e^-$-pairs from the $\rho$ vector 
meson are emitted. The thin lines denote calculation where the full branching ratio into di-leptons is attached
to both the $\rho$ decay and the $\rho$ absorption vertices. The thick lines indicate the results obtained by 
the shining method with continuous emission of di-leptons. From top to
bottom we display central Au+Au/Pb+Pb collisions at 2, 11 and 30 AGeV.}
\label{dens2}
\end{figure}

In Fig. \ref{rate} we show the respective gain and loss rates of $\rho$ mesons separated for
the different contributions for central Au+Au and Pb+Pb reactions at $E_{\rm lab}=2,11$ and $30A~$GeV. 
 The different 
processes \footnote{All processes
where the number of in-going $\rho$ mesons equals the number of
outgoing $\rho$ mesons have been discarded from the analysis as they provide only a trivial off-set.} 
in which a $\rho$ meson is produced are denoted as ``gain (collisions)'' or ``gain (decays)'' 
(meaning the stem from the decay of another resonance). The loss term
differentiates also between $\rho$ mesons which have decayed ``loss (decayed)'' and those which are absorbed in collisions
``loss (absorbed)'' as discussed above.

At $2A~$GeV, one observes that the production rate of $\rho$ mesons is dominated by the decay of 
resonances ($\sim 80$\%, full line) as compared to the formation in a s-channel $\pi\pi$ scattering 
($\sim$ 20 \%, dotted line). 
This observation is in line with the expectations that a dominant production channel for the $\rho$ 
meson in low and intermediate energy heavy ion collisions is the decay of baryon 
resonances \cite{Winckelmann:1994mx}. A previous 
detailed analysis of the $\rho$ meson production channels at $2A~$GeV within the UrQMD approach, 
found in \cite{Vogel:2005pd} confirms this interpretation in detail. The maximum of the $\rho$ meson 
production rate coincides with the maximum baryon density around $t \sim 9$~fm. 
However, these $\rho$ mesons are subject to frequent interactions with the surrounding baryons resulting
in rather short life times of the $\rho$ mesons as indicated by the large absorption rate (dashed line).
Only towards the end of the high density stage (when the $\rho$ absorption processes, 
e.g. $\rho+B \rightarrow B^{*\prime}$ cease) $\rho$ mesons can decay directly as denoted by the dashed-dotted line.
In the present model $\rho$ meson absorption accounts for the main loss 
of $\rho$ mesons, while the decay accounts for only 30\% of the $\rho$ meson loss. 
It is clear that these features  might lead to different time-dependent di-lepton 
yields, depending on the method with which the di-lepton rates are extracted from the numerical simulation.

At higher energies ($11A$ GeV, $30A$ GeV) this low energy line of arguments changes. 
Here one observes two distinct phases for the production and decay/absorption of the $\rho$ meson. 
Initially $\rho$ meson production from collisions proceeding either via string formation and fragmentation 
or via meson-meson scattering dominates the gain term (dotted line). However, also the absorption probability is 
rather high in this stage of the reaction resulting in a quick re-absorption of the $\rho$ meson (dashed line). 
For di-lepton calculations it becomes clear that only a ``shining'' 
approach has the potential to provide information on this stage, whereas the approaches which depends on the decay of the
resonance do not allow to extract this information. 
However, also at higher energies, the production of $\rho$ mesons from resonance decays in the late stage 
of the reaction is sizeable. This is evident because the sequential 
processes $NN \rightarrow B^*+X$, $B^* \rightarrow \rho +X$ will need a certain time and 
therefore triggers on later stages of the reaction. 
Thus, the $\rho$ meson production from baryon resonance decays again leads 
towards a self-triggering of $\rho$ meson production and subsequent decay at rather moderate densities, even 
at energies of $11A$ GeV and $30A$ GeV.

To gauge our assumptions for the absolute importance of the different processes the integrated $\rho$ meson
rates are shown in Fig. \ref{number}. The nomenclature is the same that has been used in Fig. \ref{rate}. 
The temporal evolution of the integrals for three values of beam energies from $E_{\rm lab}=2A~$GeV to  $E_{\rm lab}=30A~$GeV is calculated. In addition, the sum of the gain ($\Sigma_{\rm gain}$) and the sum of the loss 
terms ($\Sigma_{\rm loss}$) are displayed.  If one subtracts the sum of
 the loss terms from the sum of the gain terms one gets the yield ($n(t)$) of $\rho$ mesons which is present
 at each time step of the collision. 

As discussed above, at the lowest energy the gain via resonance decays dominates over the 
gain from direct $\rho$ meson production due to kinematical constraints. However, this behaviour 
reverses already at $E_{\rm lab}=11A~$GeV. 
At the highest energy ($E_{\rm lab}=30A~$GeV) displayed already a factor of 2.5 more $\rho$ mesons 
are produced in collisions than in decays of 
other resonances. For the loss term we observe a dominance of the absolute value of decayed 
$\rho$ mesons at $E_{\rm lab}=30A~$GeV. 
\begin{figure}[t]
\vspace*{-.5cm} %\centerline{\resizebox*{!}{0.5\textheight}
\epsfig{file=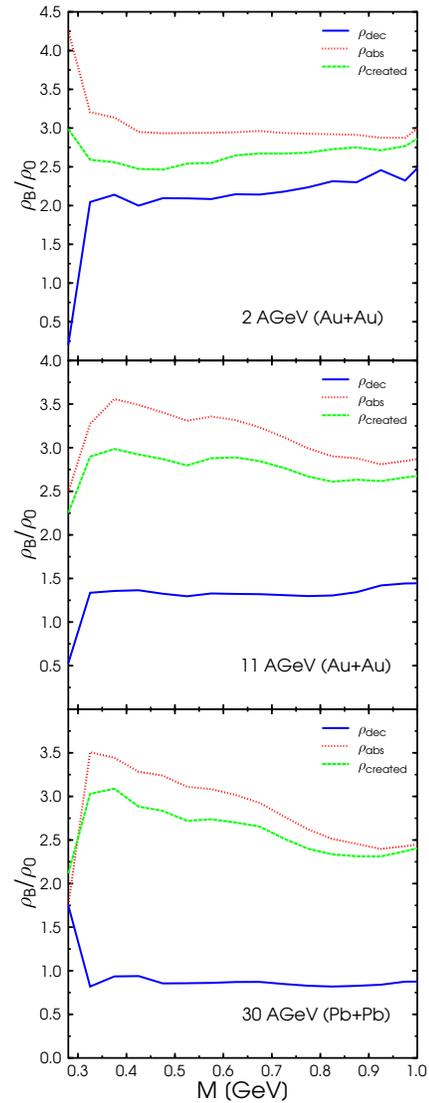,width=6cm}
\caption{(Color online) Average baryon density experienced by a $\rho$ meson as a function of the mass of the $\rho$ meson.  The results of the present calculation for the creation, absorption and decay point of the respective $\rho$ meson are shown. From top to bottom we display central Au+Au/Pb+Pb collisions at 2, 11 and 30 AGeV.}
\label{mass}
\end{figure}

Let us investigate the baryon density distribution at the space-time position of the $\rho$ meson decay and absorption 
in more detail. Fig. \ref{dens} shows the probability distribution of the baryon 
density at the instant of the $\rho$ decay (dashed line), $\rho$ absorption (dotted line) and for the 
sum of both (full line).

At $2A~$GeV (top figure) one clearly observes that absorption of the $\rho$ meson in the most dense 
medium reached at SIS energies ($\sim 3 \rho_{B}/\rho_{B0}$) is already 
a strong effect. If only decaying $\rho$ mesons are taken into account the effective density probed reduces 
to ($\sim 2 \rho_{B}/\rho_{B0}$)
At higher beam energies ($11A$~GeV, $30A$~GeV) this splitting in the density between 
decaying $\rho$ mesons and absorbed $\rho$ mesons becomes even more pronounced. 
Here, absorption processes are strongly dominating
the high baryon density stage, while decays populate the low baryon density region.

To relate the present discussion directly to the sensitivity of di-leptons to the most dense stages of the reaction, 
di-lepton rates are calculated as a function of the baryon density at which the di-leptons are 
emitted. The di-lepton calculations are based on standard cross sections as e.g. given in \cite{Schumacher:2006wc}. 
Fig. \ref{dens2} gives the  distribution of the baryon density at which the $e^+e^-$-pairs from the $\rho$ vector 
meson are emitted (from top to bottom central Au+Au/Pb+Pb collisions at 2, 11 and 30 $A$GeV are shown). 
For each energy, the di-lepton production as a function of baryon density is separated into a part where the 
initial $\rho$ meson decays (dashed lines) and a part where the $\rho$ meson was absorbed by the 
medium (dotted lines), the sum of both contributions provides the total emission rate of di-leptons and is
shown as full lines. 

Two different scenarios related to the initially discussed theoretical approaches for the di-lepton 
extraction can be  discussed:
\begin{enumerate} 
\item
In the first scenario, the full branching 
ratio into di-leptons is attached to both the $\rho$ decay and the $\rho$ absorption vertices (shown as thin lines). 
This provides the most optimistic reach towards high baryon densities, as it assumes that no collisional 
broadening takes place even in the most dense stage of the reaction (i.e. its a full vacuum baseline calculation). 
On the one hand, this setting is similar to the one employed in models that calculate di-leptons from 
folding the $\sqrt s$ distribution of nucleon-nucleon collisions as it assumes no interaction of 
the produced $\rho$ meson with the medium, i.e. absorption. 
On the other hand, if one omits the di-leptons 
emitted in the absorption process, these calculations are similar to previous calculations that assume 
di-lepton production only from the late stage decays of (baryon) resonances. 
\item
This scenario is set in contrast to the shining approach (indicated by thick lines). In the shining method, 
a continuous emission of di-leptons is assumed over the whole lifetime of the $\rho$ meson (generally for all 
relevant hadrons). The di-lepton emission rate is then integrated over time, taking the collisional broadening for each
individual vector meson in its surrounding into account. Due to the strong collisional broadening in the medium, 
a drastic reduction
in the analysis reach of di-leptons towards high densities (shown in Fig. \ref{dens2} from the comparison of 
the thick and thin full lines) results. 
At SIS and FAIR energies, the effective baryon density probed by $\rho$ mesons decaying into di-leptons is 
reduced to 1-2 $\rho_B/\rho_{B0}$ in contrast to the expected values around 2-3 $\rho_B/\rho_{B0}$ from scenario 1. 
At higher energies ($11A$~GeV, $30A$~GeV), the reach of di-leptons into the most dense stage is also strongly reduced. 
In addition, late stage decays of baryon resonances and $\rho$ mesons provide a strong trigger towards 
low baryon densities, resulting in strong peak of the di-lepton emission rate below 1-1.5 $\rho_B/\rho_{B0}$. 
This low density peak might possibly blur the view on the most interesting di-leptons from the most dense 
stages of the reaction.
\end{enumerate}

Finally, we explore the average baryon density experienced by $\rho$ mesons with different masses. This is important to understand whether any prominent features are present in the $\rho$ meson mass region between $400-600$ MeV, that is important for the intermediate mass di-lepton enhancement. 
Fig. \ref{mass} depicts the average baryon density experienced by a $\rho$ meson as a function of the mass of the $\rho$ meson. Fig. \ref{mass} gives the results of the present calculation for the creation, absorption and decay point of the respective $\rho$ meson. One observes that the baryon density of the system is constant as a function of the mass, indicating that most $\rho$ mesons, independent of their mass decay at a certain (low) baryon density, as argued before. Note that the baryon density where $\rho$ mesons are absorbed is higher, which is in line with the previous discussion.

In conclusion, we have shown that the measured di-leptons provide only a restricted view into the most dense stages 
of the reaction despite the fact that electromagnetic probes leave the reaction zone without any 
further interaction. Thus,
 possible studies of meson and baryon properties at highest baryon densities might be blurred.  
For the $\rho$ meson we have shown that the baryon density probed in the di-lepton decay channel does depend on
the method of di-lepton extraction employed. 
We argued that the absorption 
of resonances in the high baryon density region of a heavy ion collision masks 
information from the early hot and dense stage. To demonstrate this, we have split the contributions 
of the loss term of the 
$\rho$ meson yield into ``loss(absorbed)'' and ``loss(decayed)'' and have shown that at early times, i.e. at the 
highest baryon densities the absorption results in substantial reduction of the $\rho$ meson life times. 
The explored energy regime of low and intermediate energies will be investigated with the planned critRHIC program 
and is in the focus of the future FAIR project.

\begin{acknowledgments}
We thank D. Cozma and E. Bratkovskaya for useful comments.
The computational resources have been provided by the Center for the Scientific Computing (CSC) at Frankfurt. 
This work was supported by GSI, DAAD (PROCOPE) and BMBF. H.P. thanks the Deutsche Telekom Stiftung for 
financial support. H.P. and S.V. thank the Helmholtz association for additional financial support.
\end{acknowledgments}

\section*{Appendix: The four-current method}

The local baryon density at a space point $i$ is the zeroth component of the baryon four-current 
$j^\mu=(\rho_{B},\vec{j})$. The local rest frame (RF) baryon density at this space point is defined in the frame where
the three-current vanishes, $j^\mu_{RF}=(\rho_{B,RF},\vec{j}_{RF})$, with $\vec{j}_{RF}=0$. This definition is known as
the Eckart frame. Other definitions are possible, e.g. in the Landau frame, the energy-momentum tensor is at rest while 
a baryon three-current might still be present. We believe however, that the Eckart frame definition captures the 
relevant physics at the energy regime under investigation.  

In the context of the UrQMD model quantities are (per default) calculated in the 
computational frame (CF) which is (for symmetric systems) the center-of-mass frame of the whole heavy ion collision. 
In the computational frame one is only able to evaluate $j^\mu_{CF}=(\rho_{B,CF},\vec{j}_{CF})$ 
where $\rho_{B,CF}=N/V$ is
the baryon density ($N$ denoting baryon number in the volume, $V$ being the small local volume around the position $i$) 
and $\vec{j}_{CF}=\rho_{B,CF}\vec{\beta}$. In the limit of an infinitely small volume, the density $\rho_{B,CF}$ 
is a sum of Gaussians at position $i$:
\begin{eqnarray}
\rho_{CF}(\vec{r_i})&= \sum_{j=1}^N \left(\frac{1}{\sqrt{2\pi} \sigma}\right)^3 \, \gamma_{z} \,
 {\rm e}^{{\left(-\frac{(x-x_0)^2+(y-y_0)^2+(z-z_0)^2 \gamma_z^2}{2\sigma^2}\right)}}\nonumber\\
&=\sum_{j=1}^N P_j
\end{eqnarray} 
i.e., a three-dimensional in $z$-direction contracted and normalised Gaussian with 
$\gamma_z=1/\sqrt{1-\beta_z^2}$ being the Lorentz factor for the particle under consideration.
The normalisation is different for individual particles due to the different $\gamma$ factors. 
The nominal width of the Gaussian is case $\sigma=1.5$~fm. 
The particle that defines position $i$ has to be included in the sum because one is interested in the 
baryon density in the local rest frame of the cell and not in the density around a particle in its rest frame. 

The velocity of the cell is computed with the same Gaussians as used for the density 
calculation as weighting functions. Therefore, the velocity of the cell in the computational frame is: 
\begin{equation}
\vec{\beta}_{CF}=\frac{\sum_{j=1}^N \left(\frac{\vec{p_j}}{E_j}\right)\cdot P_j}{\sum_{j=1}^N P_j}\nonumber
\end{equation}
The last step is to perform a general Lorentz boost of the four-vector $j^\mu_{CF}$ into the local rest 
frame of the cell. I.e. a Lorentz transformation with the velocity of the cell $\vec{\beta_{CF}}$. 
The transformation matrix is the following:
\begin{equation}
\left(\begin{array}{cccc}
\gamma & -\beta_x \gamma & -\beta_y \gamma &-\beta_z \gamma \\
-\beta_x \gamma & 1+ (\gamma-1) \frac{\beta_x^2}{\beta^2} & (\gamma-1) \frac{\beta_x \beta_y}{\beta^2} & (\gamma-1)\frac{\beta_x\beta_z}{\beta^2}\\
-\beta_y \gamma & (\gamma-1) \frac{\beta_y \beta_x}{\beta^2} & 1+(\gamma-1) \frac{\beta_y^2}{\beta^2} & (\gamma-1)\frac{\beta_y\beta_z}{\beta^2}\\
-\beta_z \gamma & (\gamma-1) \frac{\beta_z \beta_x}{\beta^2} & (\gamma-1) \frac{\beta_z \beta_y}{\beta^2} & 1+(\gamma-1)\frac{\beta_z^2}{\beta^2}\nonumber\\
\end{array}\right)\nonumber
\label{boost}\nonumber
\end{equation}
with $\beta^2=\beta_x^2+\beta_y^2+\beta_z^2$ and $\gamma=1/\sqrt{1-\beta^2}$.
The zero-component of the transformed $j^\mu$ four-vector is the local rest frame baryon density we are interested in.

\end{document}